\documentclass{ws-procs975x65}

\def\beq{\begin{equation}}
\def\eeq{\end{equation}}

\begin{document}

\title{A novel approach to thin-shell wormholes and applications}

\author{Francisco S.~N.~Lobo} 

\address{Instituto de Astrof\'isica e Ci\^encias do Espa\c{c}o, Universidade de Lisboa, \\
	Faculdade de Ci\^encias, Campo Grande, PT1749-016 Lisboa, Portugal.
}

\author{Mariam Bouhmadi-L\'opez}
\address{Departamento de F\'{i}sica, Universidade da Beira Interior, 6200 Covilh\~a, Portugal\\
Centro de Matem\'atica e Aplica\c{c}\~oes da Universidade da Beira Interior (CMA-UBI)}

\author{Prado Mart\'{i}n-Moruno}
\address{Departamento de F\'{i}sica Te\'{o}rica I, Universidad Complutense de Madrid,\\ E-28040 Madrid,
Spain.}

\author{Nadiezhda Montelongo-Garc\'{i}a}
\address{Departamento de F\'{i}sica,
Centro de Investigaci\'{o}n y de Estudios Avanzados del IPN,
A.P. 14-740, 07000 M\'{e}xico D.F., M\'{e}xico.}

\author{Matt Visser}
\address{School of Mathematics and Statistics, Victoria University of Wellington, \\PO Box 600, Wellington 6140, New Zealand.}

\begin{abstract}
A novel framework is presented that can be adapted to a wide class of generic spherically symmetric thin-shell wormholes. By using the Darmois--Israel formalism, we analyze the stability of arbitrary spherically symmetric thin-shell wormholes to linearized perturbations around static solutions. We demonstrate in full generality that the stability of the wormhole is equivalent to choosing suitable properties for the exotic material residing on the wormhole throat. As an application, we consider the thin-shell variant of the Ellis wormhole for the cases of a vanishing momentum flux and non-zero external fluxes. 
\end{abstract}

\keywords{wormholes, thin-shell formalism, stability analysis}

\bodymatter

\section{Setting the stage}

In this paper, we report and describe an extremely general, flexible, and robust framework that can be adapted to general spherically symmetric traversable wormholes in $3+1$ dimensions. We consider standard general relativity, where all of the exotic material is confined to a thin shell.~\cite{Garcia:2011aa} Furthermore, we then analyze the stability of the thin-shell to linearized spherically symmetric perturbations around static solutions, by choosing suitable properties for the exotic material residing on the junction interface radius. Thus, in this novel approach, the stability is related to the properties of the exotic matter residing on the wormhole throat. We emphasize that the stability analysis can deal with the imposition of ``external forces'', through a momentum flux term that impinges on the shell, which is a feature that has so far been missing from the published literature.


To outline the general formalism, consider two distinct spacetime manifolds, namely, an \emph{exterior} ${\cal M_+}$, and an \emph{interior} ${\cal M_-}$, that are joined together across a surface layer $\Sigma$. Consider two generic static spherically symmetric spacetimes given by the following line elements:
\begin{eqnarray}
\hspace{-1.0cm}ds^2 = - e^{2\Phi_{\pm}(r_{\pm})}\left[1-\frac{b_{\pm}(r_{\pm})}
{r_{\pm}}\right] dt_{\pm}^2 + 
\left[1-\frac{b_{\pm}(r_{\pm})}{r_{\pm}}\right]^{-1}\,dr_{\pm}^2 + r_{\pm}^2 
d\Omega_{\pm}^{2},
\label{generalmetric}
\end{eqnarray}
where the $\pm$ signs refer to the exterior and interior geometry, respectively.

Now, relative to the surface layer, the Lanczos equations \cite{Israel} provide the following surface stress-energy components:~\cite{Garcia:2011aa}
\begin{eqnarray}
\sigma&=&-\frac{1}{4\pi a}\left[
 \sqrt{1-\frac{b_{+}(a)}{a}+\dot{a}^{2}}
+
\sqrt{1-\frac{b_{-}(a)}{a}+\dot{a}^{2}}
\right],
\label{gen-surfenergy2}
\\
{\cal P}&=&\frac{1}{8\pi a}\left[
\frac{1+\dot{a}^2+a\ddot{a}-\frac{b_+(a)+ab'_+(a)}{2a}}{\sqrt{1-\frac{b_{+}(a)}
{a}+\dot{a}^{2}}}     
+
\sqrt{1-\frac{b_{+}(a)}{a}+\dot{a}^{2}} \; a\Phi'_{+}(a)
\right. \nonumber\\
&& 
\qquad 
\left. +
\frac{1+\dot{a}^2+a\ddot{a}-\frac{b_-(a)+ab'_-(a)}{2a}}{\sqrt{1-\frac{b_{-}(a)}
{a}+\dot{a}^{2}}}
+
\sqrt{1-\frac{b_{-}(a)}{a}+\dot{a}^{2}} \; a\Phi'_{-}(a)
\right],
\label{gen-surfpressure2}
\end{eqnarray}
where $\sigma$ is the surface energy density, and ${\cal P}$ is the surface pressure, respectively. 
These results generalize those of the earlier references \citenum{Visser:1989} and \citenum{Poisson:1995},
and closely related gravastar models in reference \citenum{Visser:2003}.
The surface mass of the thin shell is given by $m_s(a)=4\pi a^2\sigma$.
The conservation equation is given by the following relation:
\begin{equation}
\frac{d(\sigma A)}{d\tau}+{\cal P}\,\frac{dA}{d\tau}=\Xi \,A \, \dot{a}\,,
\label{E:conservation3}
\end{equation}
where $A=4\pi a^2$ is the surface area of the thin shell, and the flux term is given by
\begin{equation}
\Xi =\frac{1}{4\pi a}\, \left[
\Phi_+'(a)\sqrt{1-\frac{b_+(a)}{a}+\dot{a}^{2}}
+
\Phi_-'(a)\sqrt{1-\frac{b_-(a)}{a}+\dot{a}^{2}}
\right]\,,
\end{equation}
which corresponds to the net discontinuity in the momentum flux which impinges on the shell. 

In the conservation equation (\ref{E:conservation3}), the first term represents the variation of the internal energy of the shell, the second term is the work done by the shell's internal force, and the third term represents the work done by the external forces. One can prove this equation through brute force, by explicitly differentiating equation~\eqref{gen-surfenergy2} using equation \eqref{gen-surfpressure2}. However, the conservation identity can also be obtained in a more elegant manner by taking into account the first and second contracted Gauss--Codazzi equations and the Lanczos equations.~\cite{Israel,Garcia:2011aa, 
Visser:1989, Poisson:1995, Visser:2003, Lobo:2003xd, MartinMoruno:2011rm}

Note that the vanishing of the flux term $\Xi$ is actually a quite common occurrence in the literature. But, in full generality the flux term is non-zero, and one needs the full version of the conservation equation.

\section{Linearised stability analysis}
\subsection{Equation of motion}

To analyze the stability of the thin shell, it is useful to rearrange the expression of $\sigma(a)$ into the form ${1\over2}\dot{a}^2+V(a)=0$,
where the potential $V(a)$ is given by
\begin{equation}
V(a)= {1\over2}\left\{ 1-{\bar b(a)\over a} -\left[\frac{m_{s}(a)}{2a}\right]^2-\left[\frac{\Delta(a)}{m_{s}
(a)}\right]^2\right\}\,.
   \label{potential}
\end{equation}
The quantities $\bar b(a)$ and $\Delta(a)$ are defined, for simplicity, as
\begin{eqnarray}
\bar b(a)=\frac{b_{+}(a)+b_{-}(a)}{2},\qquad
\Delta(a)=\frac{b_{+}(a)-b_{-}(a)}{2}. \nonumber
\end{eqnarray}
Note that $V(a)$ is a function of the surface mass $m_s(a)$, so that it is useful to reverse the logic flow and determine the surface mass as a function of the potential:
\begin{equation}
	m_s(a) = -a\left[ \sqrt{ 1- {b_+(a)\over a} - 2V(a)} + \sqrt{ 1- {b_-(a)\over a} - 
	2V(a)} \right]\,.
\end{equation}
Thus, if we specify $V(a)$, this tells us how much surface mass we need to put on the junction surface, which is implicitly making demands on the equation of state of the matter residing on the transition layer.

After imposing the equation of motion for the shell one has
\begin{eqnarray}
\sigma &=&-\frac{1}{4\pi a}\left[\sqrt{1-\frac{b_{+}}{a} - 2 V} + \sqrt{1-\frac{b_{-}}{a}- 2 V}\right],
\label{gen-surfenergy2-onshell} \\
 {\cal P}&=&\frac{1}{8\pi a}\left[
\frac{1-2V-aV'-\frac{b_+ +ab'_+}{2a}}{\sqrt{1-\frac{b_{+}}{a}-2V}}     
+
\sqrt{1-\frac{b_{+}}{a}-2V} \; a\Phi'_{+}
\right. \nonumber\\
&&
\qquad
\left. +
\frac{1-2V-aV'-\frac{b_- +ab'_-}{2a}}{\sqrt{1-\frac{b_{-}}{a}-2V}}
+
\sqrt{1-\frac{b_{-}}{a}-2V} \; a\Phi'_{-}
\right],
\label{gen-surfpressure2-onshell} \\
\Xi &=&\frac{1}{4\pi a}\, \left[\Phi'_+ \sqrt{1-\frac{b_+}{a}-2V} + \Phi'_- \sqrt{1-\frac{b_-}{a}-2V}\right]\,.
\end{eqnarray}
Note that the three quantities $\sigma(a)$, ${\cal P}(a)$, and $\Xi(a)$ are related by the conservation law, so only two of them are independent.

Now consider a linearization around a static solution, $a_0$, and a Taylor expansion of $V(a)$ around $a_0$ to second order. Expanding around a static solution $\dot a_0=\ddot a_0 = 0$, we have $V(a_0)=V'(a_0)=0$, so it is sufficient to consider
\begin{equation}
V(a)= \frac{1}{2}V''(a_0)(a-a_0)^2+O[(a-a_0)^3]
\,.   \label{linear-potential}
\end{equation}
The assumed static solution at $a_0$ is stable if and only if $V(a)$ has a local minimum at $a_0$, which requires $V''(a_{0})>0$. The latter condition will be our primary criterion for the thin shell stability, though it will be useful to rephrase it in terms of more basic quantities.

\subsection{Master equations}

In view of the redundancies coming from the relations $m_s(a) = 4\pi\sigma(a) a^2$ and the differential conservation law, the relevant quantities to evaluate, at the assumed stable solution $a_0$, are given by $m_s''(a)$ and $\Xi''(a)$.

The stability condition $V''(a_0)\geq0$ can be translated into an explicit inequality on $m_s''(a_0)$, given by:
\begin{eqnarray}
 m_s''(a_0) &\geq&
+{1\over4 a_0^3} 
\left\{ 
{ [b_+(a_0)- a_0 b_+'(a_0)]^2\over[1-b_+(a_0)/a_0]^{3/2}} 
+ 
{ [b_-(a_0)- a_0 b_-'(a_0)]^2\over[1-b_-(a_0)/a_0]^{3/2}}
\right\}
\nonumber\\
&& 
+{1\over2} 
\left\{ 
{b_+''(a_0)\over\sqrt{1-b_+(a_0)/a_0}} 
+
{b_-''(a_0)\over\sqrt{1-b_-(a_0)/a_0}} \right\},
  \label{stable_ddms1}
\end{eqnarray}
provided $b_+(a_0)\geq b_-(a_0)$, which is equivalent to 
$\sigma(a_0)\geq 0$.

In the absence of external forces this inequality is the only stability constraint one requires. However, once one has external forces ($\Xi\neq 0$), one obtains a second constraint, given by
\begin{eqnarray}
\left.[4\pi\,\Xi(a)\,a]''\right|_{a_0} &\leq& \left.\left\{ 
\Phi_+'''(a) \sqrt{1-b_+(a)/a} + 
\Phi_-'''(a) \sqrt{1-b_-(a)/a} \right\}\right|_{a_0}
\nonumber\\
&& 
- \left.\left\{ 
\Phi_+''(a) { (b_+(a)/a)'\over\sqrt{1-b_+(a)/a}} + 
\Phi_-''(a){(b_-(a)/a)'\over\sqrt{1-b_-(a)/a}} \right\}\right|_{a_0}
\nonumber\\
&&
-{1\over4} 
\left.\left\{ 
\Phi_+'(a) { [(b_+(a)/a)']^2\over[1-b_+(a)/a]^{3/2}} +
\Phi_-'(a) {[(b_-(a)/a)']^2\over[1-b_-(a)/a]^{3/2}} \right\}\right|_{a_0}
\nonumber\\
&&
-{1\over2} 
\left.\left\{ 
\Phi_+'(a) { (b_+(a)/a)''\over\sqrt{1-b_+(a)/a}} +
\Phi_-'(a) {(b_-(a)/a)''\over\sqrt{1-b_-(a)/a}} \right\}\right|_{a_0},
   \label{stability_Xi}
\end{eqnarray}
provided $\Phi'_{\pm}(a_0) \geq 0 $.
Note that this last quantity is entirely vacuous in the absence of external forces, which is why it has not appeared in the literature until now.~\cite{Garcia:2011aa,MartinMoruno:2011rm} 

\section{Applications: Thin-shell variant of the Ellis wormhole}

One now ``merely'' needs to apply the general formalism described above when discussing specific examples. It is perhaps instructive to consider an explicit case which violates some of the energy conditions in the bulk. Consider, for instance, the case given by the following shape functions: $b_{\pm}=R_{\pm}^2/r$, which were used in the Ellis wormhole.~\cite{ellis}

\subsection{Zero momentum flux: $\Phi_\pm =0$}

Consider the case $\Phi_\pm =0$, which implies the absence of external forces, i.e. $\Xi=0$. For this situation, ($\Phi_\pm =0$ while $b_{\pm}=R_{\pm}^2/r$), the null energy condition (NEC) is borderline satisfied in the bulk. However, the weak energy condition (WEC) is violated as the solution possesses negative energy densities, which is transparent from the following stress-energy profile: $\rho(r) = - p_r(r) = - p_t(r)=-R_\pm^2/(8\pi r^{4})$.

The  stability regions are dictated by inequality (\ref{stable_ddms1}), which yields the following dimensionless quantity
\begin{eqnarray}
a_0\,m_s''(a_0) \geq \frac{(R_+/a_0)^2}{\left( 1-R^2_+/a^2_0 \right)^{3/2}}
+
\frac{(R_-/a_0)^2}{\left( 1-R^2_-/a^2_0 \right)^{3/2}}
\,.
\label{stability_ellis}
\end{eqnarray}
It can be shown that large stability regions exist for small values of 
$R_+ \ll a_0 $ and $R_- \ll R_+ $. The stability regions decrease for
large values of $R_- \gg R_+$, and for $R_+ \gg a_0 $. (We refer the reader to reference \citenum{Garcia:2011aa} for more details.)

\subsection{Non-zero external forces: $\Phi_\pm=-R_\pm/r$}

In order to generalize the previous case of zero momentum fluxes, we now consider a case  with external forces. To this purpose, consider the 
following functions 
\begin{equation}
\Phi_\pm=-\frac{R_\pm}{r}\,.
\end{equation}
These functions imply that $\Phi'_\pm ({a_0}) > 0$, so that in addition to the stability condition given by inequality (\ref{stable_ddms1}), one needs to take into account the stability condition dictated by (\ref{stability_Xi}). The latter inequality yields the following dimensionless quantity
\begin{eqnarray}
a_0^3\,\left[4\pi a \Xi(a)\right]'' \geq 
{(6R_+/a_0-19R_+^3/a_0^3+12R_+^5/a_0^5)\over\left( 1-R^2_+/a^2_0 \right)^{3/2}}
    \nonumber \\
+{(6R_-/a_0-19R_-^3/a_0^3+12R_-^5/a_0^5)\over\left( 1-R^2_-/a^2_0 \right)^{3/2}}\,.
    \label{stability_ellis_Xi}
\end{eqnarray}
Note that as before, inequality (\ref{stable_ddms1}) yields the dimensionless quantity given by inequality (\ref{stability_ellis}). The stability regions lie above the surface where this inequality saturates. Now, in addition to the imposition of inequality (\ref{stability_ellis}), the stability regions are also restricted by condition (\ref{stability_ellis_Xi}),

Collecting the results outlined above, one may show that a large range of stability regions exist for low values of $R_+/a_0$ and of $R_-/R_+$. One also verifies the absence of the stability regions for $R_+ / a_0 \rightarrow 1$ and $R_- \gg R_+$. Once again, we refer the reader to reference \citenum{Garcia:2011aa} for specific details.

\section{Conclusions}

We have developed and described an extremely general and robust novel framework leading to the linearized stability analysis of dynamical spherically symmetric thin-shells. We have also shown that, in general, a flux term arises in the conservation law of the surface stresses. This term corresponds to an external force due to the net discontinuity in the (bulk) momentum flux which impinges on the shell. 
More specifically, we have considered the surface mass as a function of the potential, so that specifying the latter tells us how much surface mass needs to be placed on the transition layer. This procedure demonstrates in full generality that the stability of the thin shell is equivalent to choosing suitable properties for the material residing on the junction interface.

A subtlety in the cut-and-paste approach needs to be emphasized. In this work, the surface energy density is always negative for the (cut-and-paste) thin-shell traversable wormholes.~\cite{Garcia:2011aa} 
Due to the definition of the normals on the junction interface, as compared to working with gravastars, 
a few strategic sign flips arise in the context of these thin-shell traversable wormhole configurations.~\cite{Garcia:2011aa,Bouhmadi-Lopez:2014gza} Thus, the possibility of a vanishing surface energy density exists for a wide class of thin-shell configurations.~\cite{Bouhmadi-Lopez:2014gza} 
In the latter context, one can also consider an alternative approach --- by matching interior spherically symmetric wormhole solutions to an exterior vacuum geometry. Furthermore, the stability of the thin-shell to linearized spherically symmetric perturbations around static solutions was analysed. Novel wormhole solutions were found, that were supported by a matter content that minimally violates the null energy condition.~\cite{Bouhmadi-Lopez:2014gza} It is interesting to note that considering this approach, applications to gravastars have also been extensively analysed.~\cite{MartinMoruno:2011rm}

\section*{Acknowledgments}
FSNL was supported by a FCT Research contract, with reference IF/00859/2012.\\
MBL was supported by the ``Funda\c{c}\~{a}o para a Ci\^{e}ncia e Tecnologia'', through an Investigador FCT Research contract, reference IF/01442/2013/CP1196/CT0001. \\
PMM acknowledges financial support from the Spanish Ministry of Economy and Competitiveness (MINECO) through the postdoctoral training contract FPDI-2013-16161, and the project FIS2014-52837-P. \\
MV was supported by a James Cook fellowship, and by the Marsden fund, both administered by the Royal Society of New Zealand.

\end{document}